\documentclass[prd,twocolumn,preprintnumbers,amsmath,amssymb]{revtex4}
\usepackage{graphicx}
\usepackage{dcolumn}
\usepackage{bm}
\begin{document}
\title{Measurements of branching fractions for inclusive $\overline K^0/K^0$ and
$K^*(892)^{\mp}$ decays of neutral and charged $D$ mesons}
\author{
M.~Ablikim$^{1}$,              J.~Z.~Bai$^{1}$, Y.~Ban$^{12}$,
J.~G.~Bian$^{1}$,              X.~Cai$^{1}$, H.~F.~Chen$^{16}$,
H.~S.~Chen$^{1}$,              H.~X.~Chen$^{1}$, J.~C.~Chen$^{1}$,
Jin~Chen$^{1}$
% ...
\\
\vspace{0.2cm}
(BES Collaboration)}
\vspace{0.2cm} \affiliation{
\begin{minipage}{145mm}
$^{1}$ Institute of High Energy Physics, Beijing 100049, People's Republic of China\\
% ...
$^{12}$ Peking University, Beijing 100871, People's Republic of China\\
% ...
$^{16}$ University of Science and Technology of China, Hefei 230026, People's Republic of China\\
% ...
\end{minipage}
}
\email{liuhh@mail.ihep.ac.cn (H.H. Liu)}
\date{\today}

\begin{abstract}
Using the data sample of about 33 pb$^{-1}$ collected at and around
3.773 GeV with the BES-II detector at the BEPC collider, we have studied
 inclusive $\overline K^0/K^0$ and $K^*(892)^{\mp}$ decays of $D^0$ and
$D^+$ mesons. The branching fractions for the inclusive $\overline K^0/K^0$ and
$K^*(892)^-$ decays are measured to be
$BF(D^0\to \overline K^0/K^0X)=(47.6\pm4.8\pm3.0)\%$,
$BF(D^+\to \overline K^0/K^0X)=(60.5\pm5.5\pm3.3)\%$,
$BF(D^0\to K^{*-}X)=(15.3\pm 8.3\pm 1.9)\%$ and
$BF(D^+\to K^{*-}X)=(5.7\pm 5.2\pm 0.7)\%$.
The upper limits of the branching fractions for the inclusive $K^*(892)^+$
decays are set to be $BF(D^0\to K^{*+}X)<3.6\%$ and $BF(D^+\to K^{*+}X)
<20.3\%$ at 90\% confidence level.
\end{abstract}

\maketitle

\section{Introduction}
Measurements of the branching fractions for inclusive $\overline K^0/K^0$
and $K^{*\mp}$ decays of $D$ mesons are important in understanding of the 
$D$ decay mechanisms. Comparing the measured inclusive branching fraction
with the sum of those for the exclusive decays~\cite{pdg06} provides some 
information about the decay modes which have not been observed yet. 
In addition, measurements of the branching fractions for the inclusive 
$K^{*-}$ and $K^{*+}$ decays of $D$ mesons 
can also help us to study the relative strength of the Cabibbo-favored and
Cabibbo-suppressed decays. 
Up to now, these branching fractions have not been measured yet.

This Letter reports measurements of the branching fractions for 
the inclusive decays $D\to \overline K^0/K^0X$ ($X$=any particles) 
and $D\to K^{*\mp}X$.
The branching fractions are obtained based on analyses of the data 
sample of integrated luminosity of 33
pb$^{-1}$ collected with the BES-II detector at and around 3.773 GeV.
Throughout the Letter, charge conjugation is implied.

\vspace{-0.3cm}
\section{BES-II detector}
The BES-II is a conventional cylindrical magnetic detector~\cite{t1}
operated at the Beijing Electron-Positron Collider (BEPC).
A 12-layer Vertex Chamber (VC) surrounding
the beryllium beam pipe provides input to the event trigger, as well as
coordinate information. A forty-layer main drift chamber (MDC) located
just outside the VC yields precise measurements of charged
particle trajectories with a solid angle coverage of $85\%$ of 4$\pi$; it
also provides ionization energy loss ($dE/dx$) measurements which are
used for particle identification. Momentum
resolution of $1.7\%\sqrt{1+p^2}$ ($p$ in GeV/$c$) and $dE/dx$
resolution of $8.5\%$ for Bhabha scattering electrons are obtained for
the data taken at $\sqrt{s}=3.773$ GeV. An array of 48 scintillation
counters surrounding the MDC measures the time of flight (TOF) of
charged particles with a resolution of about 180 ps for electrons.
Outside the TOF, a 12 radiation length, lead-gas barrel shower counter
(BSC), operating in limited streamer mode, measures the energies of
electrons and photons over $80\%$ of the total solid angle with an
energy resolution of $\sigma_E/E=0.22/\sqrt{E}$ ($E$ in GeV) and spatial
resolutions of $\sigma_{\phi}=7.9$ mrad and $\sigma_Z=2.3$ cm for
electrons. A solenoidal magnet outside the BSC provides a 0.4 T
magnetic field in the central tracking region of the detector. Three
double-layer muon counters instrument the magnet flux return and serve
to identify muons with momentum greater than 500 MeV/c. They cover
$68\%$ of the total solid angle.

\vspace{-0.3cm}
\section{Data analysis}
Around the center-of-mass energy of 3.773 GeV, $\psi(3770)$ is produced in 
the annihilation of $e^+e^-$. 
It decays to $D\bar{D}$ pairs ($D^0\bar{D}^0$ or $D^+D^-$) with
a large branching fraction of about (85$\pm$6\%)~\cite{besbr}.
These provide us a unique method to 
directly measure the branching fractions for $D$ meson decays.  In the 
analyses 
we first reconstruct a $\bar D$ meson of the $D\bar D$ pair (this is 
called a singly tagged $\bar D$), then select the inclusive decays
$D\to \overline K^0/K^0X$ or $D \to K^{*\mp}X$ on the recoil side of the
singly tagged $\bar D$, and measure the absolute branching fractions
for these decays.

\subsection{Events selection}
To select the candidate events for the decays,  it is first required 
that at least two
charged tracks be
well reconstructed in the MDC with good helix fits. In order to ensure the
well-measured 3-momentum vectors and the reliability of the
charged-particle identification, the polar angle $\theta$ of each charged
track must satisfy $|\rm{cos\theta}|<0.85$. It is then required that each
charged track, except for those from $K^0_S$, originate from the
interaction region defined by $\sqrt{V_x^2+V_y^2}<2.0$ cm and $|V_z|<20.0$
cm, where $V_x$, $V_y$ and $V_z$ are the closest approach of the charged
track in the $x$, $y$ and $z$ directions.

Pions and kaons are identified using the $dE/dx$ and TOF measurements,
with which the combined confidence levels ($CL_{\pi}$ or $CL_{K}$) for a pion 
or kaon hypotheses are calculated. A pion candidate is required to have
$CL_{\pi}>0.001$ and a kaon candidate is required to satisfy $CL_{K}>CL_{\pi}$.

Neutral pions are reconstructed through the decay $\pi^0 \to \gamma\gamma$.
For the $\gamma$ from $\pi^0$ decay, the energy deposited in the
BSC is required to be greater than 70 MeV;
the electromagnetic shower is required to start in the first 5
readout layers; and the angle between the
$\gamma$ and the nearest charged track is required to be greater than
$22^{\circ}$~\cite{d0kev,dpk0ev}.

\subsection{Singly tagged $\bar D^0$ and $D^-$ samples}
The singly tagged $\bar D^0$ and $D^-$ samples used in the analyses have been
selected in the previous works~\cite{d0kev,dpk0ev}, where
the $\bar D^0$ mesons
are reconstructed in four hadronic decay  modes $K^+\pi^-$, $K^+\pi^-\pi^-\pi^+$,
$K^0\pi^+\pi^-$ and $K^+\pi^-\pi^0$ ($Kn\pi$, $n$= 1, 2, 3), and the 
$D^-$ mesons are reconstructed in
nine hadronic decay modes  $K^+\pi^-\pi^-$, $K^0 \pi^-$, $K^0 
K^-$, $K^+
K^-\pi^-$, $K^0\pi^-\pi^-\pi^+$, $K^0\pi^-\pi^0$, $K^+\pi^-\pi^-\pi^0$, $K^+\pi^+\pi^-
\pi^-\pi^-$ and $\pi^+\pi^-\pi^-$ ($mKn\pi$, $m$=0, 1, 2; $n$=0, 1, 2, 3, 4). 
These give the total numbers $7584\pm198(\rm stat.)\pm341(\rm sys.)$
 singly tagged $\bar D^0$ mesons~\cite{d0kev} and $5321\pm149(\rm stat.)\pm
160(\rm sys.)$ singly tagged $D^-$ mesons~\cite{dpk0ev}.

\subsection{Candidates for $D \to \overline K^0/K^0X$ and $D \to
K^{*-}(K^{*+})X$}
Candidates for the inclusive decays $D \to \overline K^0/K^0X$ and $D \to
K^{*-}(K^{*+})X$ are selected from the survival tracks on the recoil side
of the singly tagged $\bar D$. Neutral kaons are reconstructed through the
decay $K^0_S \to \pi^+\pi^-$. We require that $\pi^+\pi^-$ must originate
from a secondary vertex which is displaced from the event primary vertex 
by 7 mm at least. $K^{*-}(K^{*+})$ mesons are reconstructed through the 
decay $K^{*-}(K^{*+})\to K^0_S\pi^-(K^0_S\pi^+)$.

In each invariant masses spectrum for the $mKn\pi$ combinations, 
the region within a $\pm3\sigma_{M_{\bar D_i}}$ window around the fitted
$\bar D$ mass $M_{\bar D_i}$ is defined as the singly tagged $\bar D$ signal
region, where $\sigma_{M_{\bar D_i}}$ is the standard deviation of the
mass spectrum for the $i$th tag mode. The region outside a 
$\pm4\sigma_{M_{\bar D_i}}$ 
window around the fitted $\bar D$ mass is taken as the $\bar
D$ sideband region. In order to estimate the number of background events in
the $\bar D$ signal regions, the number of the $\bar D$ sideband events is
normalized by the ratio of the area of the fitted background in the 
$\bar D$ signal region to that of the $\bar D$ sideband.

\begin{figure}[htbp]
\begin{center}
\includegraphics[height=6cm,width=8cm]{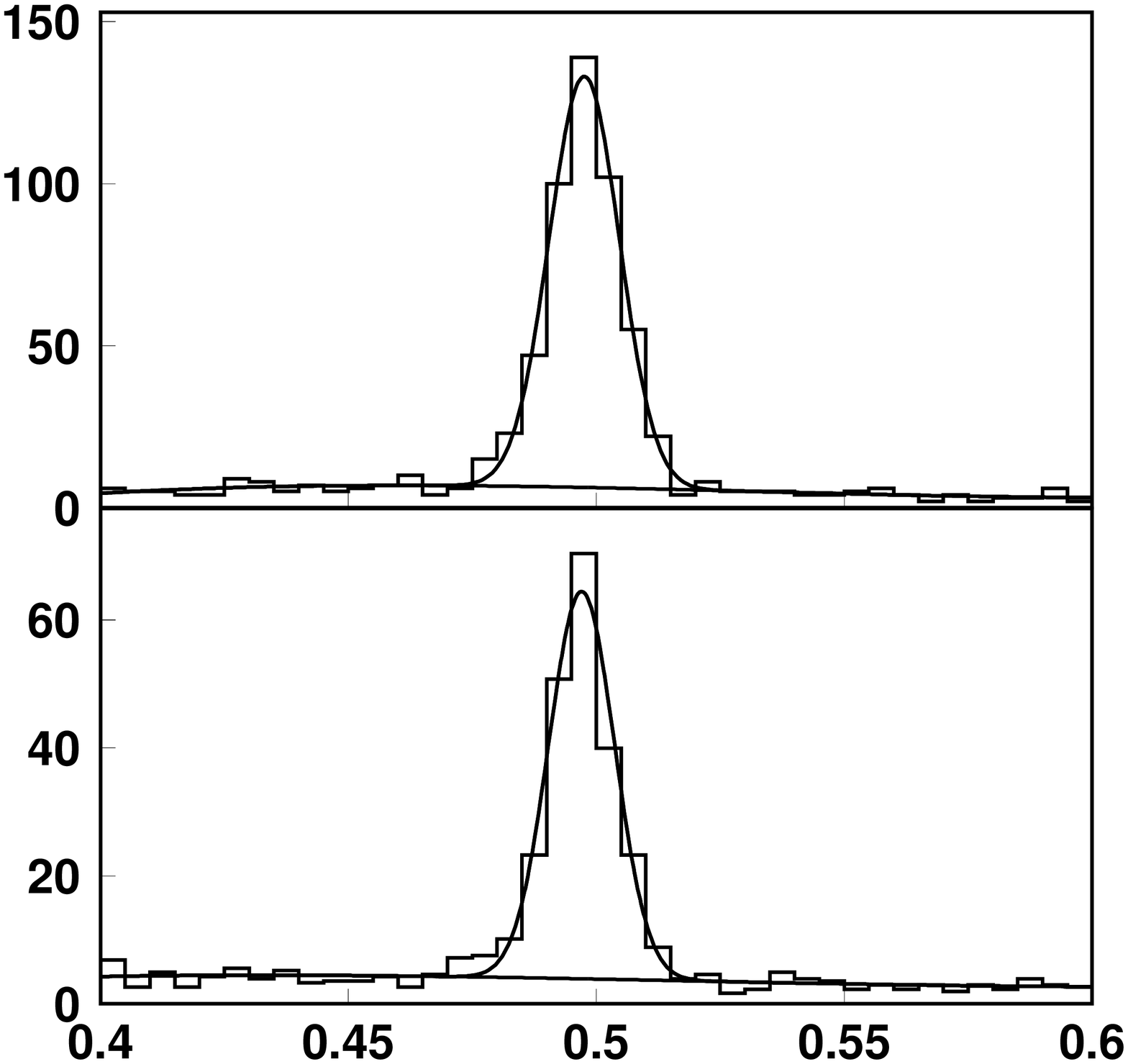}
\put(-200,-10){\bf \large Invariant Mass ($\pi^+\pi^-$) (GeV/$c^2$)}
\put(-240,15){\rotatebox{90}{\bf \large Events/(0.005 GeV/$c^2$)}}
\put(-50,140){\bf \large (a)}
\put(-50,70){\bf \large (b)}
\caption{The
distributions of the $\pi^+\pi^-$ invariant masses for the events observed
on the recoil side of the $\bar D^0$ tags for studying $D^0 \to
\overline K^0/K^0X$; (a) is the mass spectrum for the events with the
$Kn\pi$ invariant masses 
in the $\bar D^0$ signal region; (b) is the normalized mass spectrum for 
the $\bar D^0$ sideband events.}
\label{ksx_d0}
\end{center}
\end{figure}

\begin{figure}[htbp]
\begin{center}
\includegraphics[height=6cm,width=8cm]{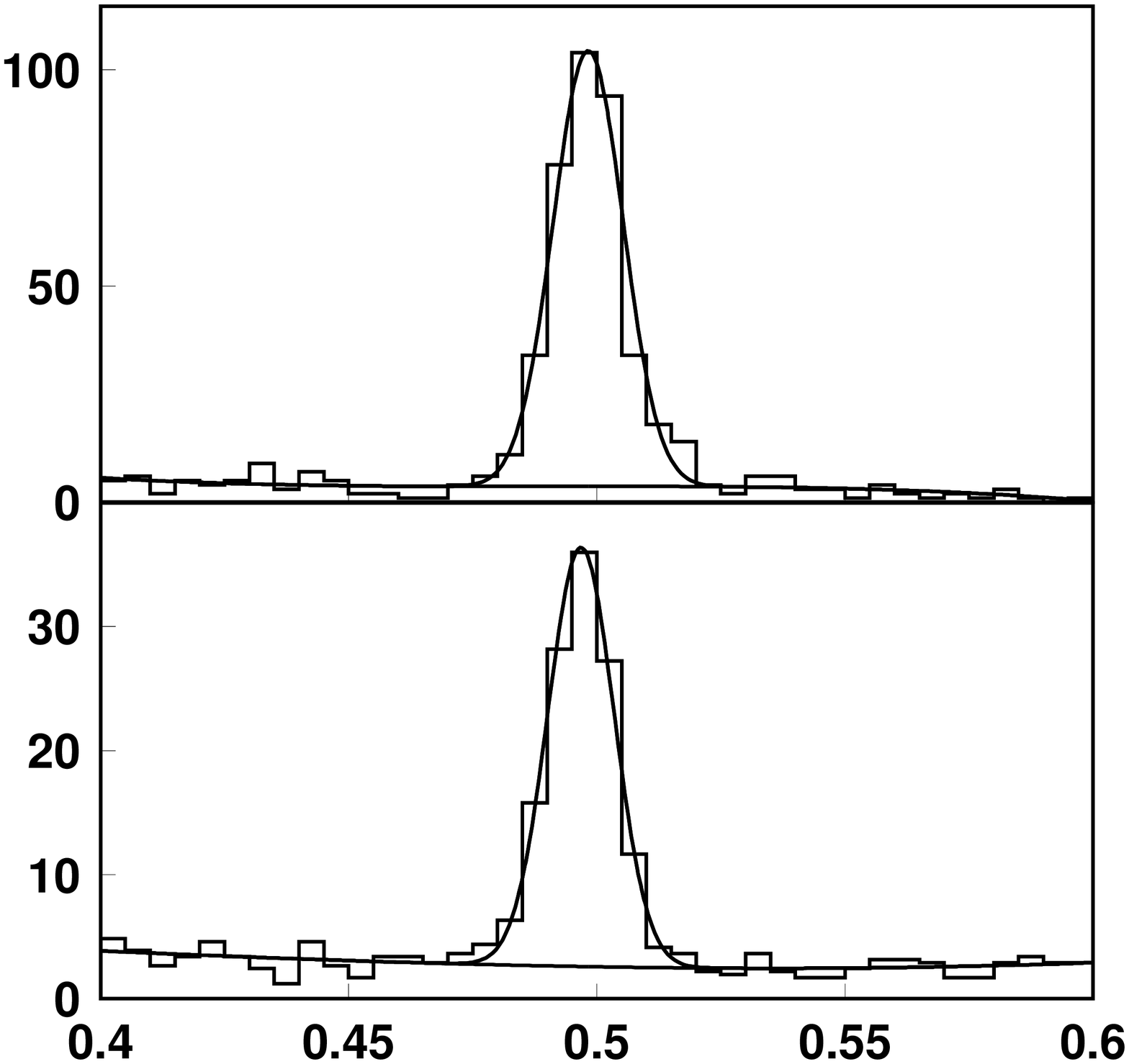}
\put(-200,-10){\bf \large Invariant Mass ($\pi^+\pi^-$) (GeV/$c^2$)}
\put(-240,15){\rotatebox{90}{\bf \large Events/(0.005 GeV/$c^2$)}}
\put(-50,140){\bf \large (a)}
\put(-50,70){\bf \large (b)}
\caption{The
distributions of the $\pi^+\pi^-$ invariant masses for the events observed
on the recoil side of the $D^-$ tags for studying $D^+ \to
\overline K^0/K^0X$; (a) is the mass spectrum for the events with the 
$mKn\pi$ invariant masses 
in the $D^-$ signal regions; (b) is the normalized mass spectrum for the
$D^-$ sideband events.}
\label{ksx_dp}
\end{center}
\end{figure}

Figures \ref{ksx_d0} and \ref{ksx_dp} show the distributions of the
$\pi^+\pi^-$ invariant masses for the events observed on the recoil side of
the $\bar D^0$ and $D^-$ tags for studying $D \to \overline K^0/K^0X$. In each
figure, (a) is the mass spectrum for the events with the $mKn\pi$
invariant masses in the $\bar D$ signal regions, and (b) is the normalized
mass spectrum for the
$\bar D$ sideband events. Fitting each mass spectrum with a Gaussian
function for $K^0_S$ signal and a polynomial to describe the background
shape, the numbers of $K^0_S$ mesons are obtained. These numbers are
summarized in Table~\ref{table1}, where $N$ and $N_b$ are the numbers of $K^0_S$
mesons observed from the events with the $mKn\pi$ invariant masses
in the $\bar D$ signal and
$\bar D$ sideband regions, respectively. Subtracting $N_b$ from $N$,
we obtain the number $n$ of the signal events for $D\to \overline
K^0/K^0X$.

\begin{figure}[htbp]
\begin{center}
\includegraphics[height=6cm,width=8cm]{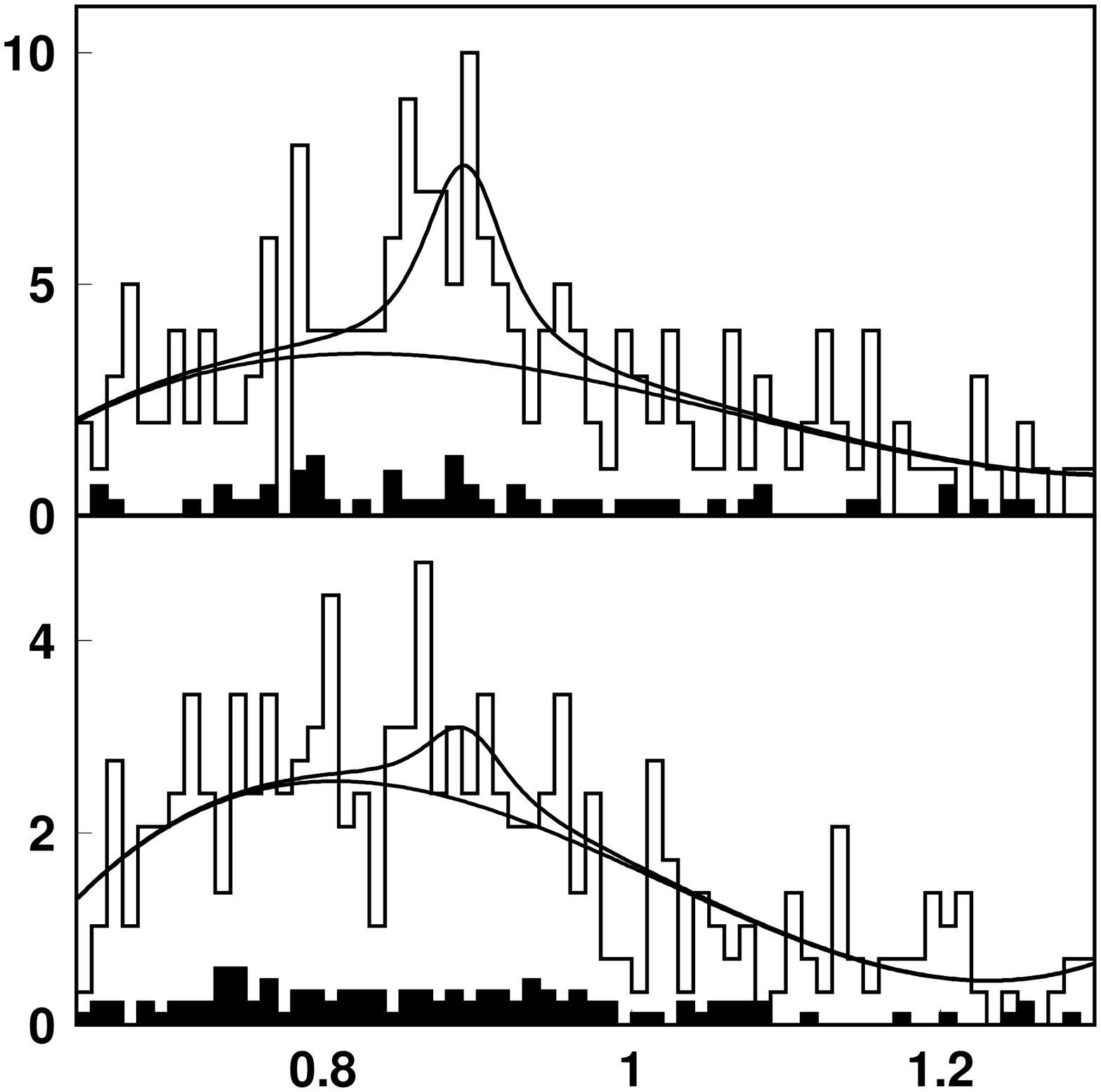}
\put(-200,-10){\bf \large Invariant Mass ($K^0_S\pi^-$) (GeV/$c^2$)}
\put(-240,15){\rotatebox{90}{\bf \large Events/(0.01 GeV/$c^2$)}}
\put(-50,140){\bf \large (a)}
\put(-50,70){\bf \large (b)}
\caption{
The
distributions of the $K^0_S\pi^-$ invariant masses for the events observed
on the recoil side of the $\bar D^0$ tags for studying $D^0 \to
K^{*-}X$; (a) is the mass spectrum for the events with the $Kn\pi$
invariant masses in the $\bar D^0$ signal regions; (b) is the normalized
mass spectrum for the $\bar D^0$ sideband events; the shadows
are the normalized background estimated by $K^0_S$ sideband.}
\label{kstar_d0}
\end{center}
\end{figure}

\begin{figure}[htbp]
\begin{center}
\includegraphics[height=6cm,width=8cm]{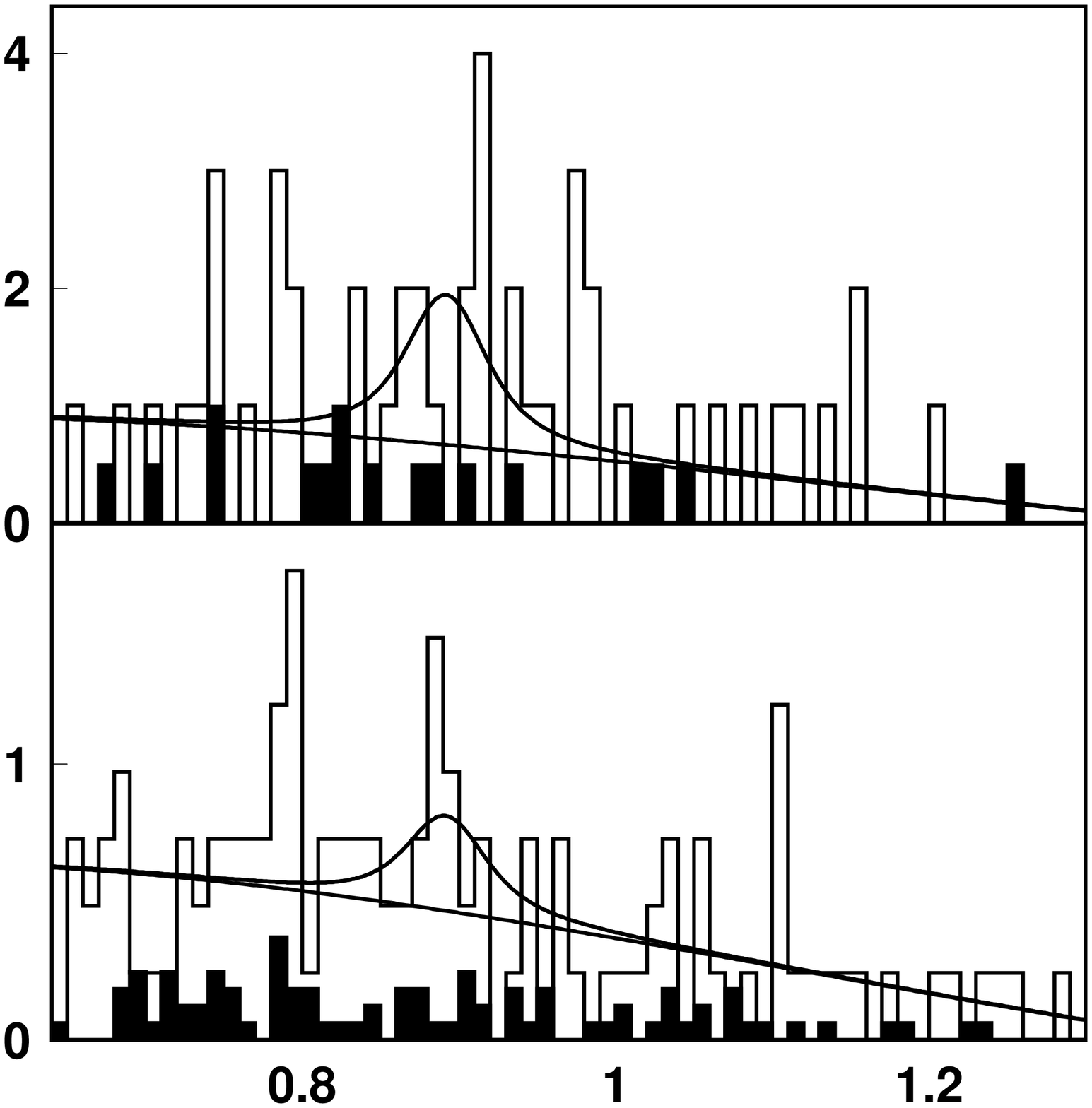}
\put(-200,-10){\bf \large Invariant Mass ($K^0_S\pi^-$) (GeV/$c^2$)}
\put(-240,15){\rotatebox{90}{\bf \large Events/(0.01 GeV/$c^2$)}}
\put(-50,140){\bf \large (a)}
\put(-50,70){\bf \large (b)}
\caption{The
distributions of the $K^0_S\pi^-$ invariant masses for the events observed
on the recoil side of the $D^-$ tags for studying $D^+ \to K^{*-}X$; 
(a) is the mass spectrum for the events with the $mKn\pi$
invariant masses in the $D^-$ signal regions; (b) is the normalized
mass spectrum for the $D^-$ sideband events; the shadows
are the normalized background estimated by $K^0_S$ sideband.}
\label{kstar_dp}
\end{center}
\end{figure}

\begin{figure}[htbp]
\begin{center}
\includegraphics[height=6cm,width=8cm]{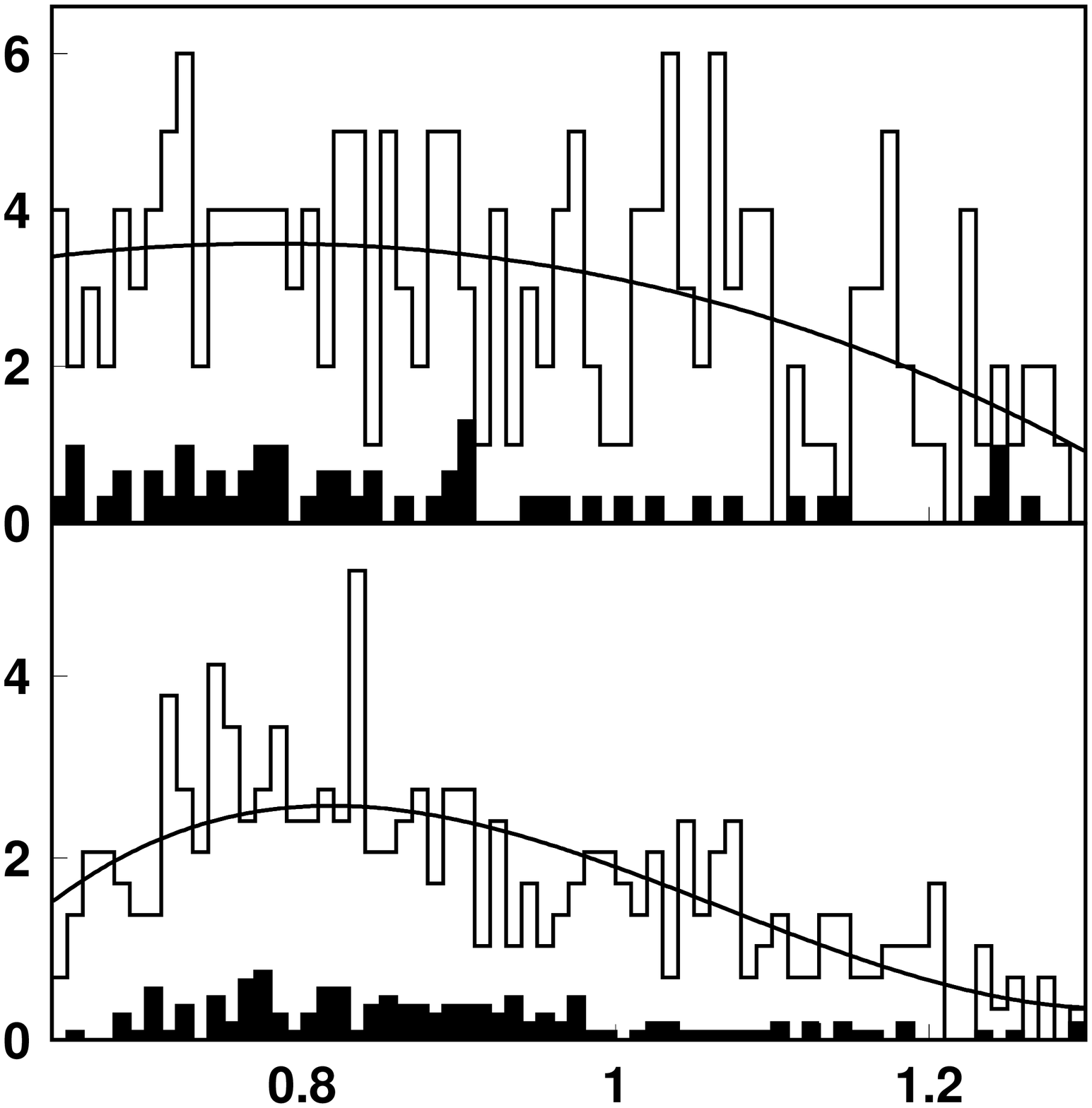}
\put(-200,-10){\bf \large Invariant Mass ($K^0_S\pi^+$) (GeV/$c^2$)}
\put(-240,15){\rotatebox{90}{\bf \large Events/(0.01 GeV/$c^2$)}}
\put(-50,140){\bf \large (a)}
\put(-50,70){\bf \large (b)}
\caption{The
distributions of invariant masses for the $K^0_S\pi^+$ 
combinations observed on the recoil side of the $\bar D^0$ tags
for studying $D^0 \to K^{*+}X$;
(a) is the mass spectrum for the events with the $Kn\pi$
invariant masses in the $\bar D^0$ signal regions; (b) is the
normalized mass spectrum for the $\bar D^0$ sideband events; the shadows
are the normalized background estimated by $K^0_S$ sideband.}
\label{kstar_d0_cbs}
\end{center}
\end{figure}

\begin{figure}[htbp]
\begin{center}
\includegraphics[height=6cm,width=8cm]{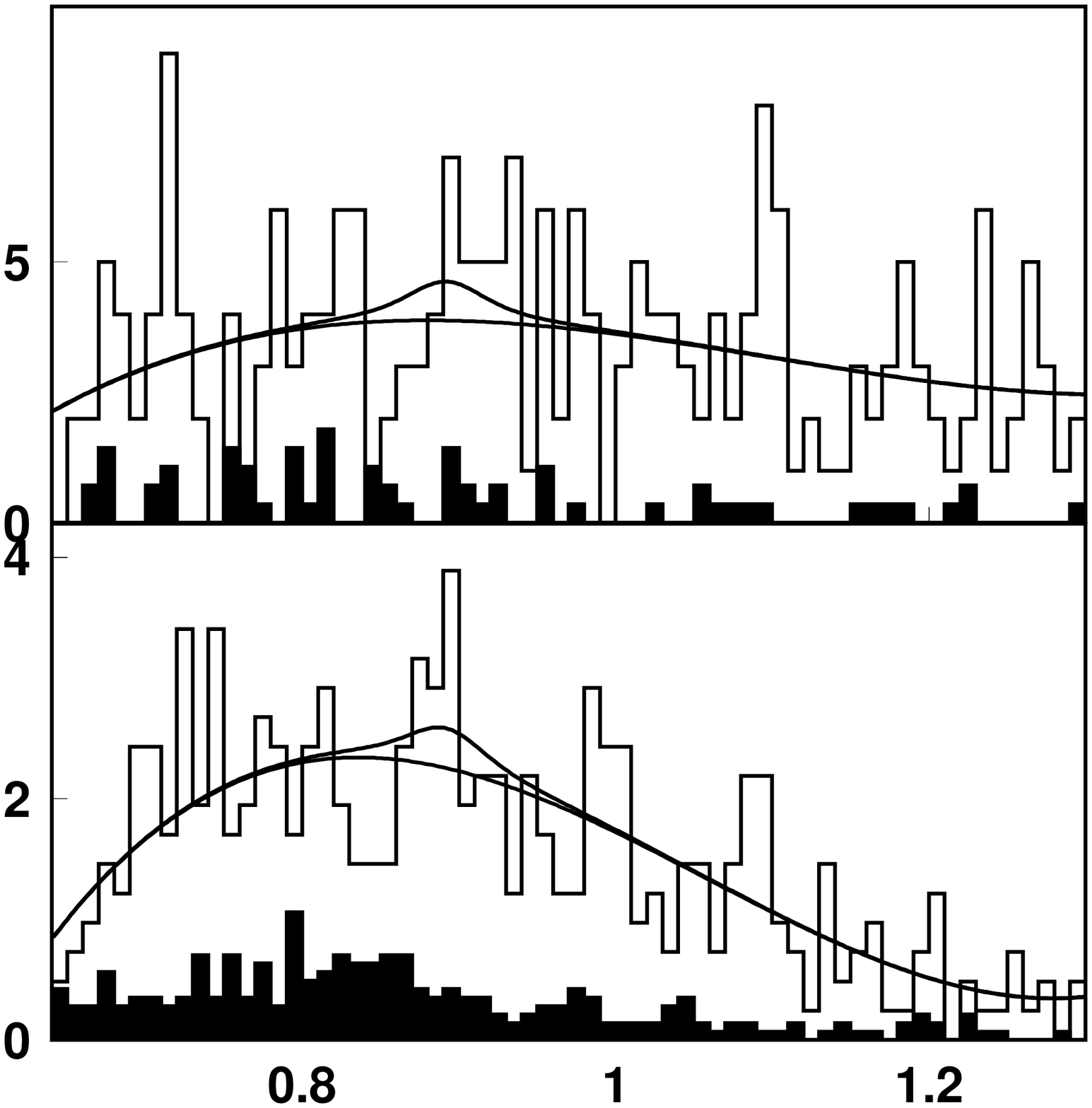}
\put(-200,-10){\bf \large Invariant Mass ($K^0_S\pi^+$) (GeV/$c^2$)}
\put(-240,15){\rotatebox{90}{\bf \large Events/(0.01 GeV/$c^2$)}}
\put(-50,140){\bf \large (a)}
\put(-50,70){\bf \large (b)}
\caption{The
distributions of the invariant masses of $K^0_S\pi^+$
combinations observed on the recoil side of the $D^-$ tags for studying
$D^+ \to K^{*+}X$; 
(a) is the mass spectrum for the events with the $mKn\pi$ invariant
masses in the $D^-$ signal regions; (b) is the normalized mass
spectrum for the $D^-$ sideband events; the shadows
are the normalized background estimated by $K^0_S$ sideband.}
\label{kstar_dp_cbs}
\end{center}
\end{figure}

Figures \ref{kstar_d0}, \ref{kstar_dp}, \ref{kstar_d0_cbs} and
\ref{kstar_dp_cbs} show the distributions of invariant masses for the 
$K^0_S\pi^-$ or $K^0_S\pi^+$ combinations observed on the recoil side of $\bar D^0$ 
or $D^-$ tags respectively for studying the decays $D \to 
K^{*-}(K^{*+})X$.
In each figure, (a) is the mass spectrum for the events in which the
$mKn\pi$ invariant masses are
in the $\bar D$ signal regions, and (b) is the normalized mass spectrum for the
$\bar D$ sideband events. The histograms are for the events with the
$\pi^+\pi^-$ invariant masses in the $K^0_S$ signal region (within
a $\pm3\sigma_{M_{K^0_S}}$ window around the fitted $K^0_S$ mass), and
the shadows are the normalized background estimated by $K^0_S$ sideband
 (outside a $\pm4\sigma_{M_{K^0_S}}$ window around the fitted $K^0_S$ mass).
Fitting each mass spectrum with a Gaussian function for $K^{*-}(K^{*+})$
signal and a polynomial to describe the background shape, we obtain the
numbers of $K^{*-}(K^{*+})$ mesons. In the fit, the mass and
width of $K^{*-}(K^{*+})$ are fixed to 0.8917 GeV/$c^{2}$ and 0.0508
GeV/$c^{2}$ quoted from PDG~\cite{pdg06}, and the detector resolution is
set to be 0.0116 GeV/$c^{2}$ determined by Monte Carlo
simulation. Table I also summarizes the numbers of $N$, $N_b$ and $n$ for
the study of $D \to K^{*-}(K^{*+})X$.

\begin{table}[htbp]
\caption{Summary of the numbers of $K^0_S$ and $K^{*\mp}$ mesons observed
on the recoil side of the $\bar D$ tags, where $N$ and $N_b$ are the 
numbers of 
$K^0_S$ or $K^{*\mp}$ mesons observed from the events
with the $mKn\pi$ invariant masses in the $\bar D$ signal regions and
$\bar D$ sideband regions, respectively, and $n$ is the number of the 
signal events of the inclusive decays.
\label{table1}
}
\begin{center}
\begin{tabular}{cccc} \hline 
Decay mode & $N$ & $N_b$ & $n$ \\ \hline
$D^0 \to \overline K^0/K^0X$ & 443.7$\pm$23.6 &
193.4$\pm$9.4 & 250.3$\pm$25.4  \\
$D^+ \to \overline K^0/K^0X$& 352.4$\pm$21.1&
108.7$\pm$6.1& 243.7$\pm$22.0    \\
$D^0 \to K^{*-}X$ & 33.7$\pm$13.8 &
6.1$\pm$5.7& 27.6$\pm$14.9 \\
$D^+ \to K^{*-}X$ & 10.0$\pm$6.2 & 2.8$\pm$2.0
&7.2$\pm$6.5 \\
$D^0 \to K^{*+}X$ & -0.7$\pm$2.8 &-0.1$\pm$2.1
&-0.6$\pm$3.5  \\
$D^+ \to K^{*+}X$ & 5.2$\pm$11.9&2.5$\pm$4.2
&2.7$\pm$12.6 \\  \hline 
\end{tabular}
\end{center}
\end{table}

\section{Results}
\subsection{Monte Carlo efficiency}
The detection efficiencies for the inclusive $\overline K^0/K^0$ and
$K^{*-}(K^{*+})$ decays of $D$ mesons are estimated by Monte Carlo
simulation. The Monte Carlo events are generated as $e^+e^-\to D\bar
D$, where $\bar D$ decays into the singly tagged $\bar D$ modes and
$D$ decays into $\overline K^0/K^0X$ or $K^{*-}(K^{*+})X$. The particle
trajectories are simulated with the GEANT3 based Monte Carlo simulation
package of the BES-II detector~\cite{t8}. Weighting the efficiencies by
the branching fractions of $D$ decays quoted from PDG~\cite{pdg06} and the 
numbers of 
the singly
tagged $\bar D$ mesons, we obtain the averaged efficiencies to be
$(6.94\pm0.06)\%$ for $D^0 \to \overline K^0/K^0X$,
$(7.57\pm0.06)\%$ for $D^+ \to \overline K^0/K^0X$,
$(2.56\pm0.04)\%$ for $D^0 \to K^{*-}(K^{*+})X$ and
$(2.36\pm0.06)\%$ for $D^+ \to K^{*-}(K^{*+})X$.

\subsection{Branching fractions}
The branching fractions for the inclusive decays $D \to \overline K^0/K^0X$
and $D \to K^{*-}X$ are determined by dividing the numbers of the signal
events by the numbers of the singly tagged $\bar D$ mesons and the
detection efficiencies. The branching fractions for the inclusive decays
are 
\begin{equation}
BF(D^0 \to \overline K^0/K^0X)=(47.6\pm4.8\pm3.0)\%,
\end{equation}
\begin{equation}
BF(D^+ \to \overline K^0/K^0X)=(60.5\pm5.5\pm3.3)\%,
\end{equation}
\begin{equation}
BF(D^0\to K^{*-}X)=(15.3\pm8.3\pm1.9)\%
\end{equation}
and
\begin{equation}
BF(D^+\to K^{*-}X)=(5.7\pm5.2\pm0.7)\%,
\end{equation}
where the first error is statistical and the second systematic.

The upper limits of the branching fractions for the inclusive decays $D^0 \to
K^{*+}X$ and $D^+ \to K^{*+}X$, which include the systematic errors, are set
to be
\begin{equation}
BF(D^0\to K^{*+}X)<3.6\%
\end{equation}
and
\begin{equation}
BF(D^+\to K^{*+}X)<20.3\%
\end{equation}
at 90\% confidence level.

In the measurement of the branching fractions for $D^0\to K^{*-}X$ 
and $D^0\to K^{*+}X$, we use three singly tagged $D^0$ modes ($K^+\pi^-$, 
$K^+\pi^-\pi^-\pi^+$ and $K^+\pi^-\pi^0$). These give us 
$7033\pm193(\rm stat.)\pm316(\rm sys.)$ singly tagged $\bar D^0$ mesons.
In the measured branching fractions, the
systematic error arises from the uncertainties in particle
identification ($\sim 0.5 \%$ per track), in tracking
($\sim 2.0\%$ per track), in the number of the singly
tagged $\bar D$ mesons ($\sim 4.5\%$ for $\bar D^0$ and
$\sim 3.0\%$ for $D^-$)~\cite{d0kev,dpk0ev}, in $K^0_S$
selection ($\sim 1.1\%$)~\cite{dpk0ev}, in background
parameterization (1.3$\% \sim 9.3\%$) and in Monte Carlo statistics
(0.8$\% \sim 2.5\%$). Adding these uncertainties in quadrature
yields the total systematic error to be
$6.4\%$ for $D^0 \to \overline K^0/K^0X$,
5.4$\%$ for $D^+ \to \overline K^0/K^0X$,
$12.2\%$ for $D^0 \to K^{*-}(K^{*+})X$ and
$11.9\%$ for $D^+ \to K^{*-}(K^{*+})X$. 
Table \ref{table2} presents the comparison of the measured branching
fractions with those measured by MARKIII~\cite{t7} and those given
by PDG~\cite{pdg06}.

\begin{table}[htbp]
\caption{Comparison of the measured branching fractions for the inclusive
decays with those measured by MARKIII~\cite{t7} and those
given by PDG~\cite{pdg06}.}
\begin{center}
\begin{tabular}{cccc} \hline 
$BF$(\%) & BES-II & MARKIII & PDG \\ \hline $D^0 \to
\overline K^0/K^0X$ & 47.6$\pm$4.8$\pm$3.0 &
45.5$\pm$5.0$\pm$3.2 & 42$\pm$5  \\
$D^+ \to \overline K^0/K^0X$ & 60.5$\pm$5.5$\pm$3.3 &
61.2$\pm$6.5$\pm$4.3 & 61$\pm$8 \\
$D^0 \to K^{*-}X$ & 15.3$\pm$8.3$\pm$1.9  &
- & - \\
$D^+ \to K^{*-}X$ & $5.7 \pm 5.2\pm 0.7$
& - &- \\
$D^0 \to K^{*+}X$ & $<$3.6 &- &-
\\
$D^+ \to K^{*+}X$ & $<$20.3 & - &-
\\  \hline 
\end{tabular}
\label{table2}
\end{center}
\end{table}

\section{Summary}
In conclusion, using the data sample of about 33 $\rm pb^{-1}$ collected at
and around 3.773 GeV with the BES-II detector at the BEPC collider, we have
studied the inclusive $\overline K^0/K^0$ and $K^{*\mp}$ decays of $D$ mesons.
The branching fractions for the inclusive $\overline K^0/K^0$ decays
are determined to be
$BF(D^0 \to \overline K^0/K^0X)=(47.6\pm 4.8\pm3.0)\%$ and
$BF(D^+ \to \overline K^0/K^0X)=(60.5\pm 5.5\pm3.3)\%$, which are in
good agreement with those measured by MARKIII and those given by PDG.
The branching fractions for the inclusive $K^{*-}$ decays are
determined to be
$BF( D^0 \to K^{*-}X)=(15.3\pm8.3\pm1.9)\%$ and
$BF( D^+ \to K^{*-}X)=(5.7\pm5.2\pm 0.7)\%$. These are measured for 
the first time. 
The upper limits of the branching fractions for the inclusive $K^{*+}$
decays are set to be
$BF(D^0 \to K^{*+}X)<3.6\%$ and
$BF(D^+ \to K^{*+}X)<20.3\%$ at 90\% confidence level.

\section{ACKNOWLEDGMENTS}
The BES collaboration thanks the staff of BEPC and computing center
for their hard efforts. This work is supported in part by the
National Natural Science Foundation of China under contracts Nos.
10491300, 10225524, 10225525, 10425523, the Chinese Academy of
Sciences under contract No. KJ 95T-03, the 100 Talents Program of
CAS under Contract Nos. U-11, U-24, U-25, and the Knowledge
Innovation Project of CAS under Contract Nos. U-602, U-34 (IHEP),
and the National Natural Science Foundation of China under Contract
No. 10225522 (Tsinghua University).

\end{document}